\documentclass[conference]{IEEEtran}
\usepackage{enumitem}
\usepackage{amsmath,amssymb,amsfonts} 
\usepackage{algorithmic}
\usepackage{textcomp}
\usepackage{xcolor}
\usepackage[utf8]{inputenc} 
\usepackage{graphicx}       
\usepackage{cite}           
\usepackage{url}            
\usepackage[hidelinks]{hyperref} 
\usepackage{float}
\usepackage{dblfloatfix}
\usepackage{xcolor}
\usepackage{booktabs}

\def\BibTeX{{\rm B\kern-.05em{\sc i\kern-.025em b}\kern-.08em
		T\kern-.1667em\lower.7ex\hbox{E}\kern-.125emX}}
\begin{document}
	
	\title{Tracing Vulnerabilities in Maven: A Study of CVE lifecycles and Dependency Networks}
	
	\author{\IEEEauthorblockN{Corey Yang-Smith}
		\IEEEauthorblockA{
			\textit{University of Calgary}\\
			Calgary, Canada \\
			corey.yangsmith@ucalgary.ca}
		\and
		\IEEEauthorblockN{Ahmad Abdellatif}
		\IEEEauthorblockA{
			\textit{University of Calgary}\\
			Calgary, Canada \\
			ahmad.abdellatif@ucalgary.ca}
	}
	
	\maketitle
	\begin{abstract}
Software ecosystems rely on centralized package registries, such as Maven, to enable code reuse and collaboration. However, the interconnected nature of these ecosystems amplifies the risks posed by security vulnerabilities in direct and transitive dependencies. While numerous studies have examined vulnerabilities in Maven and other ecosystems, there remains a gap in understanding the behavior of vulnerabilities across parent and dependent packages, and the response times of maintainers in addressing vulnerabilities. This study analyzes the lifecycle of 3,362 CVEs in Maven to uncover patterns in vulnerability mitigation and identify factors influencing at-risk packages. We conducted a comprehensive study integrating temporal analyses of CVE lifecycles, correlation analyses of GitHub repository metrics, and assessments of library maintainers' response times to patch vulnerabilities, utilizing a package dependency graph for Maven. A key finding reveals a trend in "Publish-Before-Patch" scenarios: maintainers prioritize patching severe vulnerabilities more quickly after public disclosure, reducing response time by 48.3\% from low (151 days) to critical severity (78 days). Additionally, project characteristics, such as contributor absence factor and issue activity, strongly correlate with the presence of CVEs. Leveraging tools such as the Goblin Ecosystem, OSV.dev, and OpenDigger, our findings provide insights into the practices and challenges of managing security risks in Maven.
	\end{abstract}
	
	
	\section{Introduction}
	In the field of Software Engineering, package registries like Maven (Java), npm (JavaScript), and PyPI (Python), serve as centralized hubs for sharing code. These packages often establish intricate dependency networks, where vulnerabilities propagate through the ecosystem via library imports, affecting both direct and transitive dependencies \cite{kikas2017structure, dusing2022analyzing}. Vulnerabilities compromise projects and introduce attack vectors, risking faults such as Denial of Service (DoS) or privilege escalation \cite{software_vulnerabilities_overview}; however, reporting systems such as the CVE Program \cite{cve_overview} and OSV.dev \cite{osvdev} provide frameworks for tracking and managing security risks, aiding developers in vulnerability management. According to the 2024 Stack Overflow Developer Survey \cite{stackoverflow2024survey}, 30\% of respondents reported using Java in developing software applications. As a foundational ecosystem for Java development, Maven hosts more than 47 million packages \cite{mavenrepository}, underscoring the crucial need for a deeper understanding of the systemic risks introduced through vulnerabilities.

    Prior research has focused on ecosystems like PyPI \cite{alfadel2023empirical} or npm \cite{Alfadel2022OnTD,10079181}, with some studies examining response times in Maven \cite{10172868, heng2024discovery}. However, there remains a gap in understanding the correlation between project characteristics and vulnerabilities in Maven, and the behavior of response times for package maintainers. By understanding the relationship between project characteristics and vulnerabilities, maintainers can prioritize secure and reliable dependencies, ultimately enhancing the overall quality and resilience of their projects.

    Therefore, in this study, we aim to identify key project characteristics (e.g., number of stars, issue activity) associated with vulnerabilities, analyze their severity and impact, and examine response times to better understand systemic risks within Maven. Leveraging a dependency graph of artifacts and releases hosted on Maven, sourced from the Goblin Ecosystem \cite{goblin}, we address these objectives through the following research questions:
    
	\textbf{RQ1: What is the lifecycle of vulnerabilities in Maven artifacts?} Our analysis reveals the majority of vulnerabilities (81.6\%) are patched before public disclosure, while only a small fraction (5.6\%) remain unresolved. This signals a robust effort by the community to address vulnerabilities promptly. In cases where CVE disclosure precedes patch availability (12.8\%), we observe a trend of faster response times as vulnerability severity increases.
	
    \textbf{RQ2: How do project characteristics correlate with vulnerability outcomes?} Higher issue activity, contributor absence factor, and participant counts (active contributors) correlate with the presence of vulnerable libraries. This suggests that larger, more active teams may be associated with elevated vulnerability risk.
	
	\textbf{RQ3: How long does it take for dependent packages to adopt a new fix?} We find that dependent packages most commonly adopt fixes via Available Patch Adoption cases (62.9\%), with a median of 151 days. In contrast, higher-risk Reactive Adoption (4.5\%) cases have a higher median at 249 days, often due to patch unavailability. These findings underscore the need for consistent and efficient patch adoption practices among maintainers.

\section{Methodology}
This study examines the lifecycle and contributing factors of vulnerabilities in Maven and assesses how quickly direct dependencies adopt available fixes. In this section, we outline the data and methods used to answer our research questions.

\subsection{Terminology}

\begin{itemize}
    \item \textbf{CVE Introduction Date}: Date the CVE was first introduced in a release.
    
    \item \textbf{CVE Publication Date}: Public disclosure date (per OSV.dev \cite{osvdev}).
    
    \item \textbf{CVE Patch Date}: The release date of the patched version addressing the CVE.

    \item \textbf{Latest Affected Version Date}: Last vulnerable release date before patching.
    
    \item \textbf{First Patch Adoption Date}: First release date adopting the patched version.
\end{itemize}

\subsection{CVE Lifecycle Analysis}
The Goblin Ecosystem \cite{goblin} is an open-source platform designed for ecosystem-level dependency analysis. It combines a dependency graph metamodel (Dataset), a data miner to extract dependency-related data (Miner), and a service for on-demand metrics weaving into dependency graphs (Weaver). 

The Goblin Dataset (dated 2024-08-30) is a Neo4j-based representation of the Maven ecosystem, comprised of artifacts, releases, and package interdependencies. It includes metadata such as Maven identifiers, release timestamps, versioning details, and explicitly defined dependency relationships. Using Goblin Weaver (v2.1.0) and Neo4j (Desktop Version 4.4.4), we enriched the Goblin Dataset with vulnerability information from OSV.dev \cite{osvdev}, a platform offering standardized data aligned with open-source versioning schemes and an API to retrieve publicly available vulnerability data across software ecosystems, as utilized in prior research \cite{rahman2024characterizing, zahan2022preprint}. For each CVE, we query OSV.dev to retrieve severity, vulnerable version ranges, timestamps, and alternative identifiers. We enriched the Goblin dataset with \textbf{3,362 unique CVEs} affecting 190,945 releases across 1,470 artifacts. Table \ref{tab:cve_severity} summarizes the vulnerability severity distribution. Similar to prior work \cite{alfadel2023empirical, heng2024discovery}, we define the lifecycle of a CVE as having many stages, beginning on the release date of the first vulnerable version. The lifecycle is calculated as the time delta, in days, between the CVE Introduction Date and the CVE Patch Date. To classify vulnerabilities into patch categories, we compare the chronological order between two key dates: (1) the CVE Publication Date and (2) the CVE Patch Date. A vulnerability is classified as:

\begin{itemize}
    \item \textbf{No-Patch}: No patched version exists in the dataset.
    \item \textbf{Publish-Before-Patch}: The patch is released after the vulnerability is published.
    \item \textbf{Patch-Before-Publish}: The patch is released before the vulnerability publication.
\end{itemize}

\begin{table}[t!]
    \caption{Detailed Breakdown of Severity for All CVEs, Unique}
    \begin{center}
    \begin{tabular}{l|c}
    \toprule
    \textbf{Severity} & \textbf{Count} \\ 
    \midrule
    Low               & 105 (3.12\%)   \\ 
    Medium            & 1,380 (41.05\%) \\ 
    High              & 1,279 (38.04\%) \\ 
    Critical          & 598 (17.79\%)   \\ 
    \textbf{Total}    & \textbf{3,362 (100.00\%)} \\ 
    \bottomrule
    \end{tabular}
    \label{tab:cve_severity}
    \end{center}
    \vspace{-0.40cm}
\end{table}

\subsection{Package Characteristics Correlation}

To analyze correlations between historical project characteristics and vulnerabilities, we identify GitHub as a source of metrics data that can help characterize packages we examine. Using Maven’s API \cite{sonatype_rest_api_guide}, we extracted pom.xml files from the Goblin Dataset (Section II-B), identifying 204 GitHub repositories via SCM link regex. Repository metrics (e.g., number of stars) were derived using OpenDigger \cite{Zhao_OpenDigger_2021}, which captures GHArchive \cite{gharchive} historical event data. Metrics were collected up to the CVE Patch Date (or the current date for unresolved cases) to capture the state of the system during the vulnerability period.

Of 204 repositories, 29 (14.2\%) were excluded for incomplete data (e.g., inaccessible repositories). We reduced OpenDigger’s 27 candidate repository metrics, originally sourced from CHAOSS \cite{chaoss} and X-lab \cite{xlab}, to 17 by removing those with excessive missing data (e.g., issue response time). The final dataset comprises 175 repositories relating to 456 unique CVEs, and 17 metrics spanning popularity (e.g. number of stars, forks), contributor counts, and issue activity. A detailed list of all metrics, including those excluded, and descriptions are available in our replication package.

To evaluate associations between package characteristics and the presence of CVEs, we use the \textbf{rank-biserial correlation}—a non-parametric measure of effect size for binary-continuous relationships—paired with Mann-Whitney U tests \cite{mann1947test} to determine statistical significance (p $<$ 0.05). While these tests identify meaningful associations, the severe class imbalance between historical repositories with CVEs (456) and those without (10,675) could compromise reliability.

\subsection{Dependency Patch Adoption}
To measure the response time in dependency patch adoption by package maintainers, we identify affected releases and directly dependent packages, focusing on dependencies that rely on a vulnerable parent release. For each dependent artifact, we parse its release history and query Neo4j to extract two key dates: (1) the \textbf{latest affected date}, and (2) the \textbf{first patch adoption date}. This analysis focuses on ``Publish-Before-Patch" and ``Patch-Before-Publish" cases from Section II-B, but expands to include the patch adoption behavior of dependent libraries. ``No-Patch'' cases are excluded from the analysis. Based on the order of events in CVE publication and patching, and maintainer response, we categorize dependent patch adoption behavior into three scenarios:
\begin{itemize}
    \item \textbf{Reactive Adoption:} CVE Publication $\rightarrow$ Patch Release $\rightarrow$ Dependent Adoption
    \item \textbf{Available Patch Adoption:} Patch Release $\rightarrow$ CVE Publication $\rightarrow$ Dependent Adoption
    \item \textbf{Proactive Adoption:} Patch Release $\rightarrow$ Dependent Adoption $\rightarrow$ CVE Publication
\end{itemize}
We calculate the response time for dependent packages as the time delta, in days, between the earliest patch date and the latest affected date. This allows us to evaluate the efficiency of dependency patch adoption practices and identify patterns that influence the speed of mitigation in Maven.

\section{Results}
\subsection{RQ1: What is the lifecycle of vulnerabilities in Maven artifacts?}
The distribution of vulnerabilities across the No-Patch, Publish-Before-Patch, and Patch-Before-Patch cases reveals distinct behaviors among Maven artifacts. We find that 81.6\% of CVEs in Maven are Patch-Before-Publish. Moreover, the results show that 12.8\% and 5.6\% of vulnerabilities are No-Patch and Publish-Before-Patch, respectively. Our results align with prior work \cite{heng2024discovery}, emphasizing the proactive role vendors play in managing vulnerabilities.

Table \ref{tab:mttm_severity_patchtype} presents a detailed summary of the average time taken to mitigate vulnerabilities. From the table, we observe a trend in "Publish-Before-Patch" cases: maintainers prioritize patching critical vulnerabilities more quickly after public disclosure (78 days), reflecting the heightened urgency associated with severe security issues compared to lower severity vulnerabilities (151 days). Conversely, for "Patch-Before-Publish" cases, moderate-severity issues are addressed the furthest in advance (1128 days), although no consistent trend emerges across severity levels. One potential reason for this finding is that moderate-severity issues may be detected and addressed during routine maintenance and refactoring \cite{moderate-severity-refactoring}, whereas critical-severity issues undergo the coordinated disclosure process.

\begin{table}[b!]
    \vspace{-0.40cm}
    \caption{Mean Time to Mitigate by Patch Type and Severity, in Days}
    \begin{center}
    \begin{tabular}{|c|c|c|c|c|}
    \hline
    \textbf{Patch Type} & \textbf{Low} & \textbf{Moderate} & \textbf{High} & \textbf{Critical} \\ \hline
    \textbf{Publish-Before-Patch} & 151    & 123          & 84       & 78           \\ \hline
    \textbf{Patch-Before-Publish} & $-$867 & $-$1128      & $-$897   & $-$790       \\ \hline
    \end{tabular}
    \end{center}
    \label{tab:mttm_severity_patchtype}
    \vspace{-0.40cm}
\end{table}

\subsection{RQ2: How do GitHub repository metrics correlate with project outcomes?} 

Table \ref{tab:correlation_results} presents the correlation effect size for 17 metrics sourced from OpenDigger, illustrating positive trends across all metrics (p $<$ 0.05). We observe that issue activity metrics (e.g., issue comments), repository contributor count metrics (e.g., contributor absence factor) \cite{chaoss_contributor_absence_factor} and participant count strongly correlate with the presence of CVEs, suggesting that larger, more active teams may face greater vulnerability challenges. This may be in part due to communication overhead associated with team size, which introduces complexity and reduced productivity \cite{teamsize_productivity}.

Moderate correlations were observed between vulnerability presence and characteristics such as code churn metrics (e.g., lines of code added) and popularity metrics (e.g., number of stars). This suggests that repositories with high activity and widespread usage may face challenges in vulnerability oversight. In contrast, inactive contributors showed negligible correlations, indicating limited relevance to vulnerability likelihood. However, these relationships are purely observational and do not imply causation. Unmeasured factors—such as code review practices, dependency management, or security tooling—may influence these outcomes. Our findings both align with and diverge from prior research. While some studies have reported a correlation between code churn and vulnerability presence \cite{code-churn}, others suggest that popularity metrics do not exhibit the same relationship \cite{age-and-commits, no-stars}.

\begin{table}[t!]
    \centering
    \caption{Rank Biserial Correlation Results}
    \label{tab:correlation_results}
    \begin{tabular}{l r}
        \toprule
        \textbf{Name} & \textbf{Rank Biserial Correlation} \\
        \midrule
        Issue Comments & 0.679 \\
        Issues Closed & 0.660 \\
        Contributor Absence Factor & 0.660 \\
        Issues New & 0.657 \\
        Participants & 0.651 \\
        Activity & 0.645 \\
        Technical Fork & 0.629 \\
        Attention & 0.539 \\
        New Contributors & 0.512 \\
        Change Requests & 0.464 \\
        Number of Stars & 0.474 \\
        Code Change Lines Sum & 0.452 \\
        Code Change Lines Add & 0.448 \\
        Code Change Lines Remove & 0.390 \\
        Change Requests Reviews & 0.336 \\
        Change Requests Accepted & 0.297 \\
        Inactive Contributors & 0.149 \\
        \bottomrule
    \end{tabular}
    \vspace{-0.40cm}
\end{table}

\subsection{RQ3: How long does it take for directly dependent packages to adopt a newly available fix?}

Finding the response times for direct dependencies to adopt patched versions is the first step to gaining insight about the response of the overall Maven ecosystem in addressing vulnerabilities. We observe that Available Patch Adoption cases dominate with 46,536 cases (62.9\%), but despite the immediate availability of fixes, maintainers often delay implementation, resulting in a median response time of 151 days. The highest-risk group, Reactive Adoption, comprises the fewest instances with 3,313 cases (4.5\%) yet exhibits a higher median time to resolution (249 days). The delay likely stems from the unavailability of patches during the critical period between vulnerability disclosure and adoption. While Proactive Adoption represents the best practice, this group is smaller with 24,144 cases (32.6\%) and exhibits significant variability in adoption timing, highlighting the challenges of achieving consistent and timely vulnerability mitigation.

\begin{figure}[t]
    \centering
    \includegraphics[width=\columnwidth]{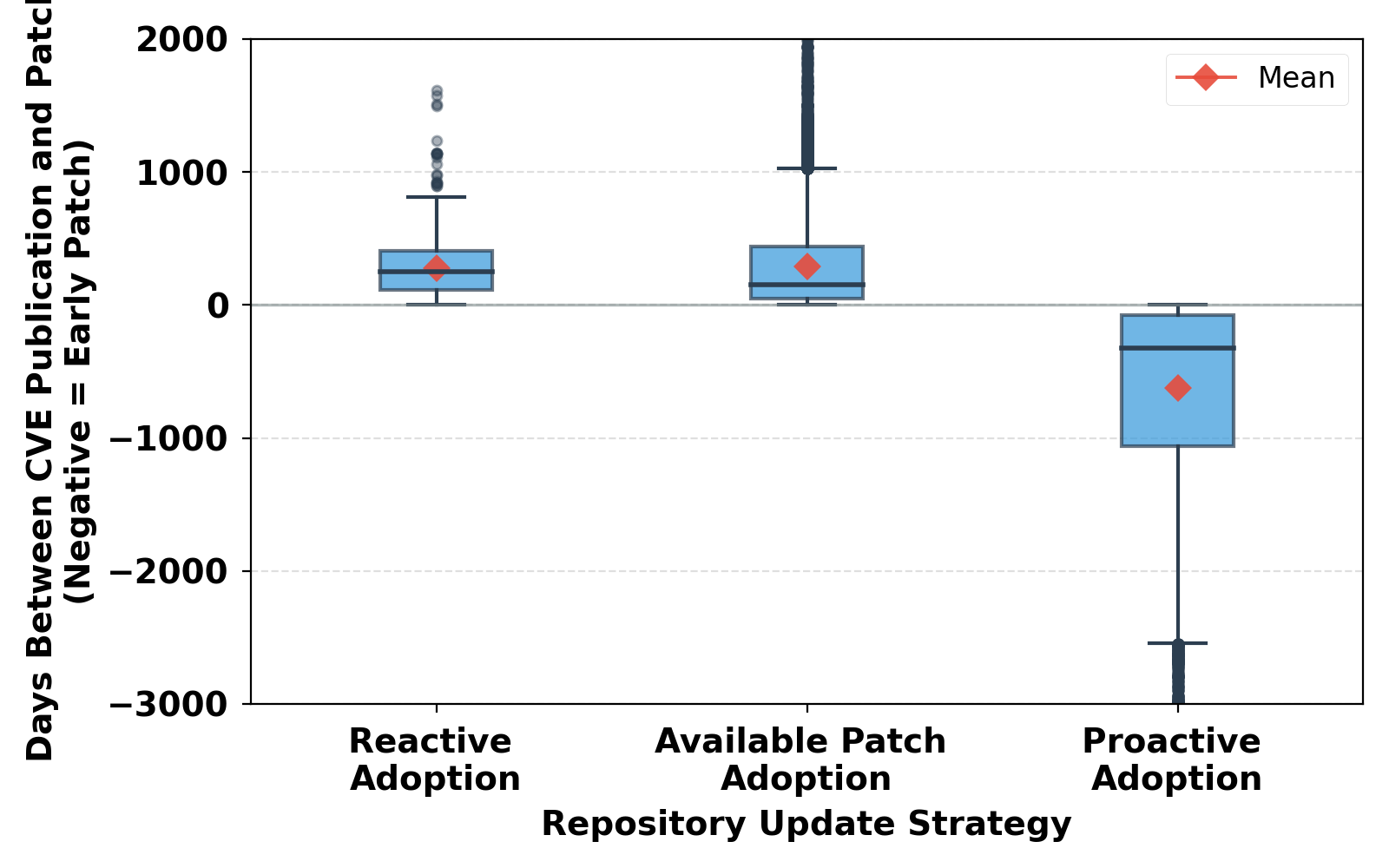}
    \vspace{-0.40cm}
    \caption{Boxplot of response time per dependent behavior.}
    \label{fig:response_times}
    \vspace{-0.40cm}
\end{figure}

Figure~\ref{fig:response_times} presents the distribution of response time behavior from directly dependent packages, and provides insights into the propagation of patches across Maven. Notably, the prevalence of Available Patch Adoption suggests that most vulnerabilities have remediations accessible to dependent packages, yet the slow median response time indicates systemic inefficiencies in prioritizing fixes. Reactive Adoption’s extended resolution period further emphasizes the risks of relying on post-disclosure patching, particularly when fixes are delayed or unavailable during critical periods. Meanwhile, the variability observed in Proactive Adoption—where some packages adopt patches rapidly while others lag—reveals unresolved challenges in achieving consistent, timely vulnerability mitigation, even among maintainers employing forward-looking practices.

\section{Threats to Validity}
\textbf{Threats to construct validity} pertain to the relationship between theoretical concepts and their empirical measurement. One key concern arises from incomplete CVE metadata, particularly the lack of explicit lower bounds for affected software versions \cite{zhang2023mitigating}. Without knowing when a vulnerability was first introduced, researchers cannot reliably assess whether older package versions remain vulnerable. This limitation may misrepresent patching behaviors in legacy systems, introducing bias and compromising the accuracy of our findings. Another challenge is the absence of CVE discovery dates. Since our calculated mean time to mitigation is based solely on observed patching behavior, it may not reflect the actual duration maintainers were aware of the vulnerability. Instead, it serves as a proxy metric, potentially underestimating true vendor response time and providing a lower bound on mitigation delays.

\textbf{Threats to internal validity} relate to potential confounding factors that may influence our results. One such factor is project size. While our analysis finds a correlation between higher issue activity and the presence of vulnerabilities, larger projects may naturally generate more issues, independent of security concerns. This raises the possibility that issue activity is a byproduct of project scale rather than a direct predictor of vulnerabilities. Future research could mitigate this confounding effect by normalizing issue activity relative to repository age, number of contributors, or total codebase size.

\textbf{Threats to external validity} concern the generalizability of our findings beyond the studied ecosystems. Our analysis focuses exclusively on the Maven ecosystem and GitHub-hosted repositories. As a result, our conclusions may not fully extend to other package registries or software configuration management (SCM) platforms. We plan (and encourage others) to replicate our methodology across diverse ecosystems and non-GitHub platforms.

\section{Related Work}
Considering the importance of mitigating and managing vulnerabilities in software ecosystems, a number of studies have been conducted in the area of vulnerability management, with efforts spanning alerting systems, Software Composition Analysis (SCA) tools, and empirical studies on vulnerability lifecycles. Cadariu et al. \cite{cadariu2015tracking} introduce a Vulnerability Alert Service for proactive risk identification, while Dietrich et al. \cite{dietrich2023security} focus on detecting code clones, addressing gaps in SCA tools.

Prior research has examined vulnerability lifecycles in PyPI \cite{alfadel2023empirical, 9647791} and npm \cite{10.1007/s10664-022-10154-1, 236368, Alfadel2022OnTD}, but fewer studies focus on Maven. For example, Heng et al. \cite{heng2024discovery} manually analyze 312 CVEs in Maven, whereas our automated approach scales to 3,362 CVEs, providing broader insights. Wu et al. \cite{10172868} explore transitive dependency risk, while we examine how direct dependents adopt security patches.

To mitigate dependency risks, Zhang et al. \cite{zhang2023mitigating} propose Ranger, an automated solution for restoring compatibility in vulnerable Maven packages. Mir et al. \cite{mir2023effecttransitivitygranularityvulnerability} use reachability analysis to assess transitive vulnerability impacts. Our work differs by empirically measuring maintainer response times, offering new insights into how security fixes propagate in Maven.

\section{Ethical Considerations}
This study raises ethical concerns about the potential misuse of the findings, particularly the correlation between larger contributor teams and increased vulnerability risks. Aligning with \cite{gold_ethics_2021}, we acknowledge that aggregated data could enable repository profiling that malicious actors might exploit. To mitigate risks, we omit repository identifiers and present results in aggregate form. By focusing on descriptive analysis rather than prescriptive measures, we limit actionable guidance that could be misapplied before vulnerabilities are resolved. Our findings aim to inform vulnerability management practices that balance ecosystem resilience with ethical responsibility.

\section{Conclusion}
This study highlights key aspects of vulnerability management in the Maven ecosystem through the analysis of 3,362 CVEs and their propagation within dependency networks. While 81.6\% of vulnerabilities are patched prior to public disclosure, demonstrating proactive efforts, vulnerabilities disclosed before a patch (12.8\%) exhibit slow response times, particularly for lower-severity issues. Furthermore, we find that project characteristics such as contributor absence factor and issue activity correlate with vulnerability presence, suggesting that larger, more active teams face increased coordination challenges. Dependency response times vary widely, with the most common approach showing a median response time of nearly five months, while the optimal strategy is both underutilized and inconsistent. These findings emphasize the need for consistent patch adoption practices, better collaboration in larger projects, and improved vulnerability management tools.

\section{Data Availability}
Our source code and dataset is available at: https://github.com/coreyyangsmith/msr2025

\bibliographystyle{IEEEtran}
\bibliography{bibliography}

\begin{thebibliography}{10}
\providecommand{\url}[1]{#1}
\csname url@samestyle\endcsname
\providecommand{\newblock}{\relax}
\providecommand{\bibinfo}[2]{#2}
\providecommand{\BIBentrySTDinterwordspacing}{\spaceskip=0pt\relax}
\providecommand{\BIBentryALTinterwordstretchfactor}{4}
\providecommand{\BIBentryALTinterwordspacing}{\spaceskip=\fontdimen2\font plus
\BIBentryALTinterwordstretchfactor\fontdimen3\font minus
  \fontdimen4\font\relax}
\providecommand{\BIBforeignlanguage}[2]{{%
\expandafter\ifx\csname l@#1\endcsname\relax
\typeout{** WARNING: IEEEtran.bst: No hyphenation pattern has been}%
\typeout{** loaded for the language `#1'. Using the pattern for}%
\typeout{** the default language instead.}%
\else
\language=\csname l@#1\endcsname
\fi
#2}}
\providecommand{\BIBdecl}{\relax}
\BIBdecl

\bibitem{kikas2017structure}
R.~Kikas, G.~Gousios, M.~Dumas, and D.~Pfahl, ``Structure and evolution of
  package dependency networks,'' in \emph{2017 IEEE/ACM 14th International
  Conference on Mining Software Repositories (MSR)}.\hskip 1em plus 0.5em minus
  0.4em\relax IEEE, 2017, pp. 102--112.

\bibitem{dusing2022analyzing}
J.~D{\"u}sing and B.~Hermann, ``Analyzing the direct and transitive impact of
  vulnerabilities onto different artifact repositories,'' \emph{Digital
  Threats: Research and Practice}, vol.~3, no.~4, pp. 1--25, 2022.

\bibitem{software_vulnerabilities_overview}
M.~C. Sánchez, J.~M.~C. de~Gea, J.~L. Fernández-Alemán, J.~Garceran, and
  A.~Toval, ``Software vulnerabilities overview: A descriptive study,''
  \emph{Tsinghua Science and Technology}, vol.~25, no.~2, pp. 270--280, 2020.

\bibitem{cve_overview}
\BIBentryALTinterwordspacing
M.~Corporation, ``Common vulnerabilities and exposures (cve): Overview,'' 2024,
  accessed: 2024-11-30. [Online]. Available:
  \url{https://www.cve.org/about/overview}
\BIBentrySTDinterwordspacing

\bibitem{osvdev}
\BIBentryALTinterwordspacing
G.~O. S.~S. Team, ``Osv: Open source vulnerabilities,'' 2024, accessed:
  2024-11-30. [Online]. Available: \url{https://osv.dev}
\BIBentrySTDinterwordspacing

\bibitem{stackoverflow2024survey}
\BIBentryALTinterwordspacing
{Stack Overflow}, ``Stack overflow developer survey 2024,'' 2024, accessed:
  2024-11-30. [Online]. Available: \url{https://survey.stackoverflow.co/2024/}
\BIBentrySTDinterwordspacing

\bibitem{mavenrepository}
\BIBentryALTinterwordspacing
``Maven repository: Central repository search,'' 2024, accessed: 2024-12-08.
  [Online]. Available: \url{https://mvnrepository.com/repos}
\BIBentrySTDinterwordspacing

\bibitem{alfadel2023empirical}
M.~Alfadel, D.~E. Costa, and E.~Shihab, ``Empirical analysis of security
  vulnerabilities in python packages,'' \emph{Empirical Software Engineering},
  vol.~28, no.~3, p.~59, 2023.

\bibitem{Alfadel2022OnTD}
\BIBentryALTinterwordspacing
M.~Alfadel, D.~E. Costa, E.~Shihab, and B.~Adams, ``On the discoverability of
  npm vulnerabilities in node.js projects,'' \emph{ACM Transactions on Software
  Engineering and Methodology}, vol.~32, pp. 1 -- 27, 2022. [Online].
  Available: \url{https://api.semanticscholar.org/CorpusID:253708713}
\BIBentrySTDinterwordspacing

\bibitem{10079181}
D.~Setó-Rey, J.~I. Santos-Martín, and C.~López-Nozal, ``Vulnerability of
  package dependency networks,'' \emph{IEEE Transactions on Network Science and
  Engineering}, vol.~10, no.~6, pp. 3396--3408, 2023.

\bibitem{10172868}
Y.~Wu, Z.~Yu, M.~Wen, Q.~Li, D.~Zou, and H.~Jin, ``Understanding the threats of
  upstream vulnerabilities to downstream projects in the maven ecosystem,'' in
  \emph{2023 IEEE/ACM 45th International Conference on Software Engineering
  (ICSE)}, 2023, pp. 1046--1058.

\bibitem{heng2024discovery}
Y.~W. Heng, Z.~Ma, H.~Zhang, Z.~Li \emph{et~al.}, ``Discovery of timeline and
  crowd reaction of software vulnerability disclosures,'' \emph{arXiv preprint
  arXiv:2411.07480}, 2024.

\bibitem{goblin}
D.~Jaime, J.~El~Haddad, and P.~Poizat, ``Navigating and exploring software
  dependency graphs using goblin,'' in \emph{Proceedings of the International
  Conference on Mining Software Repositories (MSR 2025)}, 2025.

\bibitem{rahman2024characterizing}
I.~Rahman, N.~Zahan, S.~Magill, W.~Enck, and L.~Williams, ``Characterizing
  dependency update practice of npm, pypi and cargo packages,'' \emph{arXiv
  preprint arXiv:2403.17382}, 2024.

\bibitem{zahan2022preprint}
N.~Zahan, P.~Kanakiya, B.~Hambleton, S.~Shohan, and L.~Williams, ``Preprint:
  Can the openssf scorecard be used to measure the security posture of npm and
  pypi?'' \emph{arXiv preprint arXiv:2208.03412}, 2022.

\bibitem{sonatype_rest_api_guide}
{Sonatype, Inc.}, ``Sonatype central search rest api guide,''
  \url{https://central.sonatype.org/search/rest-api-guide/}, n.d., accessed:
  2024-11-30.

\bibitem{Zhao_OpenDigger_2021}
\BIBentryALTinterwordspacing
S.~Zhao, X.~Zhang, and X.~Xia, ``{OpenDigger},'' Apr. 2021. [Online].
  Available: \url{https://github.com/X-lab2017/open-digger}
\BIBentrySTDinterwordspacing

\bibitem{gharchive}
{GH Archive}, ``{GH Archive},'' \url{http://www.gharchive.org}, 2024, accessed:
  2024-11-30.

\bibitem{chaoss}
\BIBentryALTinterwordspacing
{CHAOSS Community}, ``{CHAOSS: Community Health Analytics Open Source
  Software},'' 2025, accessed: 2025-02-05. [Online]. Available:
  \url{https://chaoss.community}
\BIBentrySTDinterwordspacing

\bibitem{xlab}
\BIBentryALTinterwordspacing
{X-Lab}, ``{X-Lab: Unlocking Innovation},'' 2025, accessed: 2025-02-05.
  [Online]. Available: \url{https://www.x-lab.info}
\BIBentrySTDinterwordspacing

\bibitem{mann1947test}
H.~B. Mann and D.~R. Whitney, ``On a test of whether one of two random
  variables is stochastically larger than the other,'' \emph{Annals of
  Mathematical Statistics}, vol.~18, no.~1, pp. 50--60, March 1947.

\bibitem{moderate-severity-refactoring}
\BIBentryALTinterwordspacing
A.~Ikegami, R.~G. Kula, B.~Chinthanet, V.~Maeprasart, A.~Ouni, T.~Ishio, and
  K.~Matsumoto, ``On the use of refactoring in security vulnerability fixes: An
  exploratory study on maven libraries,'' in \emph{Proceedings of the 26th
  International Conference on Evaluation and Assessment in Software
  Engineering}, ser. EASE '22.\hskip 1em plus 0.5em minus 0.4em\relax New York,
  NY, USA: Association for Computing Machinery, 2022, p. 288–293. [Online].
  Available: \url{https://doi.org/10.1145/3530019.3535304}
\BIBentrySTDinterwordspacing

\bibitem{chaoss_contributor_absence_factor}
\BIBentryALTinterwordspacing
{CHAOSS Project}, ``Contributor absence factor,'' 2025, accessed: 2025-02-01.
  [Online]. Available:
  \url{https://chaoss.community/kb/metric-contributor-absence-factor/}
\BIBentrySTDinterwordspacing

\bibitem{teamsize_productivity}
D.~Rodriguez, M.~Sicilia, E.~Barriocanal, and R.~Harrison, ``Empirical findings
  on team size and productivity in software development,'' \emph{Journal of
  Systems and Software - JSS}, vol.~85, 03 2012.

\bibitem{code-churn}
Y.~Shin, A.~Meneely, L.~Williams, and J.~A. Osborne, ``Evaluating complexity,
  code churn, and developer activity metrics as indicators of software
  vulnerabilities,'' \emph{IEEE Transactions on Software Engineering}, vol.~37,
  no.~6, pp. 772--787, 2011.

\bibitem{age-and-commits}
\BIBentryALTinterwordspacing
H.~Al-Shammare, N.~Al-Otaiby, M.~Al-Otabi, and M.~Alshayeb, ``An empirical
  investigation of the security weaknesses in open-source projects,'' in
  \emph{Proceedings of the 28th International Conference on Evaluation and
  Assessment in Software Engineering}, ser. EASE '24.\hskip 1em plus 0.5em
  minus 0.4em\relax New York, NY, USA: Association for Computing Machinery,
  2024, p. 634–642. [Online]. Available:
  \url{https://doi.org/10.1145/3661167.3661279}
\BIBentrySTDinterwordspacing

\bibitem{no-stars}
\BIBentryALTinterwordspacing
M.~S. Naveed, ``Correlation between github stars and code vulnerabilities,''
  \emph{Journal of Computing \& Biomedical Informatics}, vol.~4, no.~01, pp.
  141--151, Dec. 2022. [Online]. Available:
  \url{https://jcbi.org/index.php/Main/article/view/111}
\BIBentrySTDinterwordspacing

\bibitem{zhang2023mitigating}
L.~Zhang, C.~Liu, S.~Chen, Z.~Xu, L.~Fan, L.~Zhao, Y.~Zhang, and Y.~Liu,
  ``Mitigating persistence of open-source vulnerabilities in maven ecosystem,''
  in \emph{2023 38th IEEE/ACM International Conference on Automated Software
  Engineering (ASE)}.\hskip 1em plus 0.5em minus 0.4em\relax IEEE, 2023, pp.
  191--203.

\bibitem{cadariu2015tracking}
M.~Cadariu, E.~Bouwers, J.~Visser, and A.~Van~Deursen, ``Tracking known
  security vulnerabilities in proprietary software systems,'' in \emph{2015
  IEEE 22nd International Conference on Software Analysis, Evolution, and
  Reengineering (SANER)}.\hskip 1em plus 0.5em minus 0.4em\relax IEEE, 2015,
  pp. 516--519.

\bibitem{dietrich2023security}
J.~Dietrich, S.~Rasheed, A.~Jordan, and T.~White, ``On the security blind spots
  of software composition analysis,'' in \emph{Proceedings of the 2024 Workshop
  on Software Supply Chain Offensive Research and Ecosystem Defenses}, 2023,
  pp. 77--87.

\bibitem{9647791}
J.~Ruohonen, K.~Hjerppe, and K.~Rindell, ``A large-scale security-oriented
  static analysis of python packages in pypi,'' in \emph{2021 18th
  International Conference on Privacy, Security and Trust (PST)}, 2021, pp.
  1--10.

\bibitem{10.1007/s10664-022-10154-1}
\BIBentryALTinterwordspacing
A.~Zerouali, T.~Mens, A.~Decan, and C.~De~Roover, ``On the impact of security
  vulnerabilities in the npm and rubygems dependency networks,''
  \emph{Empirical Softw. Engg.}, vol.~27, no.~5, Sep. 2022. [Online].
  Available: \url{https://doi.org/10.1007/s10664-022-10154-1}
\BIBentrySTDinterwordspacing

\bibitem{236368}
\BIBentryALTinterwordspacing
M.~Zimmermann, C.-A. Staicu, C.~Tenny, and M.~Pradel, ``Small world with high
  risks: A study of security threats in the npm ecosystem,'' in \emph{28th
  USENIX Security Symposium (USENIX Security 19)}.\hskip 1em plus 0.5em minus
  0.4em\relax Santa Clara, CA: USENIX Association, Aug. 2019, pp. 995--1010.
  [Online]. Available:
  \url{https://www.usenix.org/conference/usenixsecurity19/presentation/zimmerman}
\BIBentrySTDinterwordspacing

\bibitem{mir2023effecttransitivitygranularityvulnerability}
\BIBentryALTinterwordspacing
A.~M. Mir, M.~Keshani, and S.~Proksch, ``On the effect of transitivity and
  granularity on vulnerability propagation in the maven ecosystem,'' 2023.
  [Online]. Available: \url{https://arxiv.org/abs/2301.07972}
\BIBentrySTDinterwordspacing

\bibitem{gold_ethics_2021}
\BIBentryALTinterwordspacing
N.~E. Gold and J.~Krinke, ``Ethics in the mining of software repositories,''
  \emph{Empirical Software Engineering}, vol.~27, no.~1, p.~17, Nov. 2021.
  [Online]. Available: \url{https://doi.org/10.1007/s10664-021-10057-7}
\BIBentrySTDinterwordspacing

\end{thebibliography}

\end{document}